\documentclass{ifacconf}

\usepackage{graphicx}      
\usepackage{natbib}        

\usepackage{amsmath,amsfonts}

\newcommand{\R}{\mathbb{R}}
\newcommand{\setX}{\mathcal{X}}
\newcommand{\setY}{\mathcal{Y}}
\newcommand{\SSQ}{SMC$^2$}
\newcommand{\Nt}{{N_\theta}}
\newcommand{\OO}{\mathcal{O}}
\usepackage{color}

\begin{document}
\begin{frontmatter}

\title{Towards automatic calibration of the number of state particles within the
\SSQ~algorithm\thanksref{footnoteinfo}} 

\thanks[footnoteinfo]{The first author is partially supported by a grant from
the French National Research Agency (ANR) as part of the “Investissements d'Avenir”
program (ANR-11-LABEX-0047).
The third author is supported by DARPA under Grant No. FA8750-14-2-0117. 
}

\author[First]{N. Chopin} 
\author[Second]{J. Ridgway}
\author[Third]{M. Gerber} 
\author[Fourth]{O. Papaspiliopoulos}

\address[First]{CREST-ENSAE, Malakoff, and HEC Paris,
France (e-mail: nicolas.chopin@ensae.fr)}
\address[Second]{CREST-ENSAE, Malakoff,and Universite Dauphine France (e-mail: james.ridgway@ensae.fr)}
\address[Third]{Harvard University, USA (e-mail: mathieugerber@fas.harvard.edu)}
\address[Fourth]{ICREA \& Universitat Pompeu Fabra, Barcelona, Spain (e-mail:omiros.papaspiliopoulos@upf.edu)}

\begin{abstract}                
\SSQ~\citep{smc2} is an efficient algorithm for sequential estimation
and state inference of state-space models. It generates $N_{\theta}$
parameter particles $\theta^{m}$, and, for each $\theta^{m}$, it
runs a particle filter of size $N_{x}$ (i.e. at each time step, $N_{x}$
particles are generated in the state space $\setX$). We discuss how
to automatically calibrate $N_{x}$ in the course of the algorithm.
Our approach relies on conditional Sequential Monte Carlo
  updates, monitoring the state of the pseudo random number generator
and  on an estimator of the variance of the unbiased estimate of the
likelihood that is produced by the particle filters, which is obtained
using nonparametric regression techniques. We observe that our approach 
is both less CPU intensive and with smaller Monte Carlo errors than the initial
version of \SSQ. 

\end{abstract}


\end{frontmatter}

\section{Introduction}

Consider a state-space model, with parameter $\theta\in\Theta$, latent
Markov process $(x_{t})_{t\geq0}$, and observed process $(y_{t})_{t\geq0}$,
taking values respectively in $\setX$ and $\setY$. The model is
defined through the following probability densities: $\theta$ has
prior $p(\theta)$, $(x_{t})_{t\geq0}$ has initial law $\mu_{\theta}(x_{0})$
and Markov transition $f_{\theta}^{X}(x_{t}|x_{t-1})$, and the $y_{t}$'s
are conditionally independent, given the $x_{t}$'s, with density
$f_{\theta}^{Y}(y_{t}|x_{t})$. Sequential analysis of such a model
amounts to computing recursively (in $t$) the posterior distributions
\begin{multline*}
p(\theta,x_{0:t}|y_{0:t})= \\ 
\frac{p(\theta)\mu_{\theta}(x_{0})}{p(y_{0:t})}
 \left\{ \prod_{s=1}^{t}f_{\theta}^{X}(x_{s}|x_{s-1})\right\}
\left\{ \prod_{s=0}^{t}f_{\theta}^{Y}(y_{s}|x_{s})\right\} 
\end{multline*}
or some of its marginals (e.g. $p(\theta|y_{0:t})$); the normalising constant $p(y_{0:t})$ 
 of the above density is the marginal likelihood (evidence) of the data observed up to time $t$. 

For a fixed $\theta$, the standard approach to sequential analysis
of state-space models is particle filtering: one propagates $N_{x}$ particles
in $\setX$ over time through mutation steps (based on proposal distribution $q_{t,\theta}(x_t|x_{t-1})$ at time $t$)
and resampling steps; see Algorithm \ref{alg:PF}. 
Note the conventions: $1:N_x$ denotes the set of integers $\{1,\ldots,N_x\}$, 
$y_{0:t}$ is $(y_0,\ldots,y_t)$,
$x_t^{1:N_x}=(x_t^1,\ldots,x_t^{N_x})$,
$x_{0:t}^{1:N_x}=(x_0^{1:N_x},\ldots,x_t^{1:N_x})$, and so on.

\begin{algorithm}

\protect\caption{Particle filter (PF, for fixed $\theta$)}\label{alg:PF}
\hrulefill

Operations involving superscript $n$ must be performed for all $n\in1:N_{x}$. 

At time $0$: 
\begin{description}
\item [{(a)}] Sample $x_{0}^{n}\sim q_{0,\theta}(x_{0})$.
\item [{(b)}] Compute weights 
\[
w_{0,\theta}(x_{0}^{n})=\frac{\mu_{\theta}(x_{0}^{n})f^{Y}(y_{0}|x_{0}^{n})}{q_{0,\theta}(x_{0}^{n})}
\]
normalised weights, $W_{0,\theta}^{n}=w_{0,\theta}(x_{0}^{n})/\sum_{i=1}^{N_{x}}w_{0,\theta}(x_{0}^{i})$,
and incremental likelihood estimate \\ $\hat{\ell}_0(\theta)=N_{x}^{-1}\sum_{n=1}^{N_{x}}w_{0,\theta}^n$. 
\end{description}
Recursively, from time $t=1$ to time $t=T$:
\begin{description}
\item [{(a)}] Sample $a_{t}^{n}\sim\mathcal{M}(W_{t-1,\theta}^{1:N_{x}})$,
the multinomial distribution which generates value $i\in1:N_{x}$
with probability $W_{t-1,\theta}^{i}$. 
\item [{(b)}] Sample $x_{t}^{n}\sim q_{t,\theta}(\cdot|x_{t-1}^{a_{t}^{n}})$. 
\item [{(c)}] Compute weights 
\begin{equation*}
\begin{split}
w_{t,\theta}(x_{t-1}^{a_t^n},x_t^n) & =
\frac{f^{X}(x_{t}^{n}|x_{t-1}^{a_{t}^{n}})f^{Y}(y_{t}|x_{t}^{n})}{q_{t,\theta}(x_{t}^{n}|x_{t-1}^{a_{t}^{n}})} 
\\  
W_{t,\theta}^{n} & = {w_{t,\theta}(x_{t-1}^{a_t^n},x_t^n) \over
  \sum_{i=1}^{N_{x}}w_{t,\theta}(x_{t-1}^{a_t^i},x_t^i)}
\end{split}
\end{equation*}
and incremental likelihood estimate \\ 
$\hat{\ell}_t(\theta)=N_x^{-1}\sum_{n=1}^{N_x}w_{t,\theta}(x_{t-1}^{a_t^n},x_t^n)$. 
\end{description}
\hrulefill
\end{algorithm}

The output of Algorithm \ref{alg:PF} may be used in different ways: at time $t$, the 
quantity 
$\sum_{n=1}^{N_x} W_{t,\theta}^n \varphi(x_t^n)$ is a consistent (as $N_x\rightarrow +\infty$) estimator 
of the filtering expectation $\mathbb{E} [\varphi(x_t)|y_{0:t},\theta]$; In addition, $\hat{\ell}_t(\theta)$
is an \emph{unbiased} estimator of incremental likelihood $p(y_t|y_{0:t-1},\theta)$, and 
$\prod_{s=0}^t \hat{\ell}(\theta)$ is an unbiased estimator of the full likelihood $p(y_{0:t}|\theta)$
\citep[][Lemma 3]{DelMoral1996unbiased}.

In order to perform joint inference on parameter $\theta$ and state variables, \cite{smc2} derived 
the \SSQ~sampler, that is, a SMC (Sequential Monte Carlo) algorithm in $\theta-$space, which generates
and propagates $N_\theta$ values $\theta^m$ in $\Theta$, and which, for each $\theta^m$, runs a particle
filter (i.e. Algorithm \ref{alg:PF}) for $\theta=\theta^m$, of size $N_x$. One issue however is how to choose $N_x$:
if too big, then CPU time is wasted, while if taken too small, then the performance of the algorithm deteriorates. 
 \cite{smc2} give formal results (adapted from \cite{pmcmc}) that suggest that $N_x$ should grow at a linear rate
during the course of the algorithm. They also propose a practical method for increasing $N_x$ adaptively, based 
on an importance sampling step where the $\Nt$ particle systems, of size $N_x$, are replaced by new particle
systems of size $N_x^\mathrm{new}$. But this importance sampling step increases the degeneracy of the weights, 
which in return may leads to more frequent resampling steps, which are expensive. 
In this paper, we derive an alternative way to increase $N_x$ adaptively, which is not based on importance 
sampling, but rather on a CSMC (conditional Sequential Monte Carlo) update, which is less CPU intensive.

\section{Background on \SSQ}\label{sec:background}

\subsection{IBIS}

To explain \SSQ, we first recall the structure of the IBIS algorithm \citep{Chopin:IBIS} as Algorithm \ref{alg:ibis}. 
For a model with parameter $\theta\in\Theta$, prior $p(\theta)$, data $y_{0:T}$, and incremental likelihood  $p(y_t|y_{0:t-1},\theta)$, 
IBIS provides at each iteration $t$ an approximation of partial posterior $p(\theta|y_{0:t})$. In practice, IBIS samples 
$N_\theta$ particles $\theta^m$ from the prior, then perfoms sequential importance sampling steps, from $p(\theta|y_{0:t-1})$
to $p(\theta|y_{0:t})$ using incremental weight $p(\theta|y_{0:t})/p(\theta|y_{0:t-1})\propto p(y_t|y_{0:t-1},\theta)$. 

\begin{algorithm}[h]
\protect\caption{IBIS}\label{alg:ibis}
\hrulefill 
 
Operations involving superscript $m$ must be performed for all
$m\in1:N_{\theta}$. 

\begin{description}
\item [{(Init)}] Sample $\theta^{m}\sim p(\theta)$, set $\omega^{m}\leftarrow 1$. 
\end{description}
From time $t=0$ to time $t=T$, do 
\begin{description}
\item [(a)] Update importance weights
$$ \omega^m \leftarrow \omega^m \times p(y_t|y_{0:t-1},\theta). $$ 
\item [(b)] If ESS($\omega^{1:N_{\theta}})\leq\mathrm{ESS}_{\min}$,
sample (for all $m$) $\tilde{\theta}^m$ from mixture 
$$ \frac{1}{\sum_{m=1}^\Nt \omega^m} \sum_{m=1}^\Nt  \omega^m K_t(\theta^m,d\theta),$$ 
where $K_t$ is a Markov kernel with invariant distribution $p(\theta|y_{0:t})$; 
finally reset particle system to 
$$ \theta^{1:\Nt} \leftarrow \tilde{\theta}^{1:\Nt},\quad \omega^{1:\Nt} \leftarrow (1,\ldots,1).$$
\end{description}
\hrulefill
\end{algorithm}

To avoid weight degeneracy, one performs a resample-move step (described as Step (b) in Algorithm \ref{alg:ibis}). When the 
ESS (effective sample size) of the weights, computed as: 
$$ \mathrm{ESS}(\omega^{1:\Nt})= \frac{(\sum_{m=1}^{\Nt} \omega^m)^2}{\sum_{m=1}^{\Nt} (\omega^m)^2} \in [1,N] $$
goes below some threshold $\mathrm{ESS}_{\min}$ (e.g. $N/2$), the $\theta^m$'s  are resampled, then moved according to 
some Markov kernel $K_t$ that leaves invariant the current target of the algorithm, $p(\theta|y_{0:t})$. 
This resample-move step re-introduces diversity among the $\theta$-particles. 

A convenient default choice for $K_t$ is several iterations of random-walk Metropolis, with the random step calibrated
to the spread of the current particle population (i.e. variance of random step equals some fraction of the covariance matrix
of the resampled particles). 

The main limitation of IBIS is that it requires evaluating the likelihood increment $p(y_t|y_{0:t-1},\theta)$, which is 
typically intractable for state-space models.
On the other hand, we have seen that this quantity may be estimated 
unbiasedly by particle filtering. This suggests combining IBIS (i.e. SMC in the $\theta$-dimension) with 
particle filtering (i.e. SMC in the $x_t-$dimension), as done in the \SSQ~algorithm.

\subsection{\SSQ}

The general structure of \SSQ~is recalled as Algorithm \ref{alg:smc2}. Essentially, one recognises the IBIS
algorithm, where the intractable incremental weight $p(y_t|y_{0:t-1},\theta^m)$ has been replaced by 
the unbiased estimate $\hat{\ell}_t(\theta^m)$. This estimate is obtained from a PF run for $\theta=\theta^m$; 
thus $N_\theta$ PFs are run in parallel. Denote $(x_{0:t}^{1:N_x,m},a_{1:t}^{1:N_x,m})$ the random variables
generated by the PF associated to $\theta^m$. 

\begin{algorithm}[h]
\protect\caption{\SSQ}\label{alg:smc2}
\hrulefill

Operations involving superscript $m$ must be performed for all
$m\in1:N_{\theta}$. 
\begin{description}
\item [{(Init)}] Sample $\theta^{m}\sim p(\theta)$, set $\omega^{m}\leftarrow 1$. 
\end{description}
From time $t=0$ to time $t=T$, do 
\begin{description}
\item [{(a)}] For each $\theta^{m}$, run iteration $t$ of Algorithm \ref{alg:PF},
so as to obtain $(x_{0:t}^{1:N_x,m},a_{1:t}^{1:N_x,m})$, and $\hat{\ell}_t(\theta^m)$.  
\item [{(b)}] Update weights
\[
\omega^{m}\leftarrow\omega^{m}\times\hat{\ell}_t(\theta^{m}).
\]

\item [{(c)}] If ESS($\omega^{1:N_{\theta}})\leq\mathrm{ESS}_{\min}$,
sample (for all $m$) 
$(\tilde{\theta}^m,\tilde{x}_{0:t}^{1:N_x,m},\tilde{a}_{1:t}^{1:N_x,m})$ 
from mixture 
$$ \frac{1}{\sum_{m=1}^\Nt \omega^m} \sum_{m=1}^\Nt  \omega^m K_t\left( (\theta^m,x_{0:t}^{1:N_x,m},a_{1:t}^{1:N_x,m}),d\cdot\right),$$ 
where $K_t$ is a PMCMC kernel with invariant distribution $\pi_t(\theta,x_{0:t}^{1:N_x},a_{1:t}^{1:N_x})$ (see text); 
finally reset particle system to 
$$ (\theta^m,x_{0:t}^{1:N_x,m},a_{1:t}^{1:N_x,m}) 
\leftarrow (\tilde{\theta}^m,\tilde{x}_{0:t}^{1:N_x,m},\tilde{a}_{1:t}^{1:N_x,m}) $$
and $\omega^m\leftarrow 1$, for all $m$.  

\end{description}
\hrulefill
\end{algorithm}

This `double-layer' structure suggests that \SSQ~suffers from two levels of approximation, 
and as such that it requires both $N_x\rightarrow +\infty$ and $\Nt\rightarrow +\infty$ 
to converge. It turns out however that  \SSQ~ is valid for any fixed value of $N_x$; that is,
for any fixed $N_x\geq 1$, it converges as  $\Nt\rightarrow +\infty$. 

This property is intuitive in the simplified case when resampling-move steps are never triggered
(i.e. take $\mathrm{ESS}_{\min}=0$). Then \SSQ~collapses to importance sampling, with weights replaced 
by unbiased estimates, and it is easy to show convergence from first principles. 

We now give a brief outline of the formal justification of \SSQ~for fixed $N_x$, and refer to 
\cite{smc2} for more details. \SSQ~may be formalised as a SMC sampler for the sequence of extended
distributions: 
$$
\pi_t(\theta,x_{0:t}^{1:N_x},a_{1:t}^{1:N_x}) = \frac{p(\theta)}{p(y_{0:t})}
\psi_{t,\theta}(x_{0:t}^{1:N_x},a_{1:t}^{1:N_x}) \prod_{s=0}^t \hat{\ell}_s(\theta)
$$
where $\psi_{t,\theta}$ denotes the joint pdf of the random variables generated by a PF up to time $t$ (for parameter $\theta$), 
and $\hat{\ell}_s(\theta)$ denotes the unbiased estimate of the likelihood increment computed from that PF, 
$\hat{\ell}_0(\theta)=N_x^{-1}\sum_{n=1}^N w_0(x_0^n)$,  
$\hat{\ell}_s(\theta)=N_x^{-1}\sum_{n=1}^N w_{s,\theta}(x_{s-1}^{a_t^n},x_s^n)$ for $s>0$; i.e. $\hat{\ell}_s(\theta)$
is actually a function of $(\theta,x_{0:s}^{1:N_x},a_{1:s}^{1:N_x})$.

One recognises in $\pi_t$ the type of extended target distribution simulated by PMCMC (Particle MCMC, \cite{pmcmc}) 
algorithms. Note $\pi_t$ is a proper probability density (it integrates to one), and that 
the marginal distribution of $\theta$ is $p(\theta|y_{0:t})$. These two properties are easily deduced from 
the unbiasedness of $\prod_{s=0}^t \hat{\ell}_s(\theta)$ (as an estimator of $p(y_{0:t}|\theta)$). In addition, 
\begin{multline*}
\pi_t(\theta,x_{0:t}^{1:N_x},a_{1:t}^{1:N_x})  = \\
\pi_{t-1}(\theta,x_{0:t-1}^{1:N_x},a_{1:t-1}^{1:N_x})
\frac{\psi_{t,\theta}(x_{0:t}^{1:N_x},a_{1:t}^{1:N_x})}{\psi_{t-1,\theta}(x_{0:t-1}^{1:N_x},a_{1:t-1}^{1:N_x})}
\hat{\ell}_t(\theta) 
\end{multline*}
where one recognises in the second factor the distribution of the variables generated by a PF at time $t$, conditional
on those variables generated up to time $t-1$. Thus, the equation above justifies both Step (a) of Algorithm \ref{alg:smc2}, where the particle 
filters are extended from time $t-1$ to $t$, and Step (b),
where the particles $(\theta^m,x_{0:t}^{1:N_x,m},a_{1:t}^{1:N_x,m})$ 
are reweighted by $\hat{\ell}_t(\theta^m)$.

We describe in the following section PMCMC moves that may be used in Step (c). Before, we note that a naive implementation 
of \SSQ~has a $\OO(tN_x\Nt)$ memory cost at time $t$, as one must stores in memory 
$(\theta^m,x_{0:t}^{1:N_x,m},a_{1:t}^{1:N_x,m})$ for each $m \in 1:\Nt$. This memory cost may be substantial even
on a modern computer. 

\subsection{PMCMC moves}

To make more explicit the dependence of the unbiased estimate of the likelihood on the variables generated during 
the course of PF, define
\begin{multline*}
L_t(\theta,x_{0:t}^{1:N_x},a_{1:t}^{1:N_x}) =\prod_{s=0}^t \hat{\ell}_s(\theta) \\
= \left\{\frac{1}{N_x} \sum_{n=1}^{N_x} w_{0,\theta}(x_0^n)\right\}
\prod_{s=1}^t \left\{\frac{1}{N_x} \sum_{n=1}^{N_x} w_{s,\theta}(x_{s-1}^{a_s^n},x_s^n) \right\}.
\end{multline*}

The PMMH (Particle Markov Metropolis-Hastings) kernel, described as Algorithm \ref{alg:pmmh}, 
may be described informally as a Metropolis step in $\theta$-space, where the likelihood of 
both the current value and the proposed value have been replaced by unbiased estimators. Formally,
as proven in \cite{pmcmc}, it is in fact a standard Metropolis step with respect to the extended distribution
$\pi_t(\theta,x_{0:t}^{1:N_x},a_{1:t}^{1:N_x})$; in particular it leaves invariant $p(\theta|y_{0:t})$. 
(For convenience, our description of PMMH assumes a random walk proposal, but PMMH is not restricted 
to this kind of proposal.)

\begin{algorithm}[h]
\protect\caption{Random walk PMMH update}\label{alg:pmmh}
\hrulefill

\textbf{Input:} $(\theta,x_{0:t}^{1:N_x},a_{1:t}^{1:N_x})$

\textbf{Output:} $(\tilde{\theta},\tilde{x}_{0:t}^{1:N_x},\tilde{a}_{1:t}^{1:N_x})$ 
\begin{description}
 \item[1.] $\theta^\star = \theta+z$, $z\sim N(0,\Sigma_t).$ 
 \item[2.] Generate PF (Algorithm \ref{alg:PF}) for parameter $\theta^\star$; let 
 $(x_{0:t}^{1:N_x,\star},a_{1:t}^{1:N_x,\star})$ the output. 
 \item[3.] With probability $1\wedge r$, 
 $$ r = \frac{p(\theta^\star)L_t(\theta^\star,x_{0:t}^{1:N_x,\star},a_{1:t}^{1:N_x,\star}) }
 {p(\theta)L_t(\theta,x_{0:t}^{1:N_x},a_{1:t}^{1:N_x}) } $$ 
 let 
 $(\tilde{\theta},\tilde{x}_{0:t}^{1:N_x},\tilde{a}_{1:t}^{1:N_x})
\leftarrow (\theta^{\star},x_{0:t}^{1:N_x,\star},a_{1:t}^{1:N_x,\star})$; otherwise 
 $(\tilde{\theta},\tilde{x}_{0:t}^{1:N_x},\tilde{a}_{1:t}^{1:N_x})
\leftarrow (\theta,x_{0:t}^{1:N_x},a_{1:t}^{1:N_x})$. 
\end{description}

\hrulefill
\end{algorithm}

In practice, we set $\Sigma_t$, the covariance matrix of the proposal, to a fraction of
the covariance matrix of the resampled $\theta$-particles. 

One advantage of using PMHMH within \SSQ~is that it does not require storing all the variables
generated by the $\Nt$ PFs: operations at time $t>0$ require only having access to, for each $m$, 
$(\theta^m,x_{t-1}^{1:N_x,m},a_{t-1}^{1:N_x,m})$ and $L_{t-1}(\theta^m,x_{0:t-1}^{1:N_x,m},a_{1:t}^{1:N_x,m})$,
which is computed recursively. Memory cost then reduces to $\OO(N_\theta N_x)$.

The Particle Gibbs approach is an alternative PMCMC step, based on the following property 
of target $\pi_t$: if one extends $\pi_t$ with random index $k$, such that $k\in 1:N_x$, and 
$k\sim \mathcal{M}(W_T^{1:N_x})$, the normalised weighs at the final iteration, then 
(a) the selected trajectory, together with $\theta$, follow the posterior distribution 
$p(\theta,x_{0:t}|y_{0:t})$; and (b) the remaining arguments of $\pi_t$ follow a CSMC 
(conditional SMC) distribution, which corresponds to the distribution of the random 
variables generated by a PF, but conditional on one trajectory fixed to the selected trajectory; 
see Algorithm \ref{alg:pgibbs}.

\begin{algorithm}[h]
\protect\caption{Particle Gibbs update}\label{alg:pgibbs}
\hrulefill 

\textbf{Input:} $(\theta,x_{0:t}^{1:N_x},a_{1:t}^{1:N_x})$

\textbf{Output:} $(\tilde{\theta},\tilde{x}_{0:t}^{1:N_x},\tilde{a}_{1:t}^{1:N_x})$ 
\begin{description}
 \item[1.] Sample $b_t\sim \mathcal{M}(W_t^{1:N_x})$, with 
 $W_t^n = w_{t,\theta}(x_{t-1}^{a_t^n},x_t^n)/\sum_{i=1}^{N_x} w_{t,\theta}(x_{t-1}^{a_t^i},x_t^i).$
 From $s=t-1$ to $s=0$, set $b_s \leftarrow a_{s+1}^{b_{s+1}}$. 
  Set $\tilde{x}_s^1 \leftarrow x_s^{b_s}$, $\tilde{a}_s^1=1$ for all $s \in 0:T$.
 \item[2.] Sample $\tilde{\theta}$ from a MCMC step that leaves invariant distribution 
 $ p(\theta | x_{0:t},y_{0:t}) $, but with $x_{0:t}$ set to $\tilde{x}_{0:t}^1$. 
 \item[3.] Sample $(\tilde{x}_{0:t}^{2:N_x},\tilde{a}_{1:t}^{2:N_x})$ as in 
 Algorithm \ref{alg:PF}, but for parameter $\tilde{\theta}$ and conditionally on $\tilde{x}_{0:t}^1$, that is: 
 at time $0$, generate $\tilde{x}_0^n\sim q_{0,\tilde{\theta}}$  for $n\in 2:N$, at time $1$, 
 sample $a_t^n\sim \mathcal{M}(W_1^{1:N_x})$, for $n\in 2:N$, and 
 $x_t^n \sim q_{1,\tilde{\theta}}(\cdot|\tilde{x}_{t-1}^{\tilde{a}_1^n})$, and so on. 
\end{description}

\hrulefill

\end{algorithm}

In contrast with PMMH, implementing particle Gibbs steps within \SSQ~requires having access 
to all the variables $(\theta^m,x_{0:t}^{1:N_x,m},a_{1:t}^{1:N_x,m})$ at time $t$, which as
we have already discussed, might incur too big a memory cost.

\subsection{Choosing $N_x$}

\cite{pmcmc} show that, in order to obtain reasonable performance for PMMH, one should
take $N_x=\OO(t)$. \cite{AndrieuLeeVihola} show a similar result for
Particle Gibbs.

In the context of \SSQ, this suggests that $N_x$ should be allowed to increase in the course
of the algorithm. To that effect, \cite{smc2} devised an exchange step, which consists 
in exchanging the current particle systems, of size $N_x$, with new particle systems, of size 
$N_x^{\mathrm{new}}$, through importance sampling. In \cite{smc2}'s implementation, 
the exchange step is triggered each time the acceptance rate of the PMMH step (as performed in 
Step 3. of Algorithm \ref{alg:pmmh}) 
is below a certain threshold, and $N_x^{\mathrm{new}}=2N_x$ (i.e. $N_x$ doubles every time). 

The main drawback of this approach is that it introduces some weight degeneracy immediately 
after the resampling step. In particular, we will observe in our simulations that this 
prevents us from changing $N_x$ too frequently, as the ESS of the weights then becomes too low. 

In this paper, we discuss how to use a Particle Gibbs step in order to increase $N_x$ without changing the 
weights.

\section{Proposed approach}

\subsection{Particle Gibbs and memory cost}

We first remark that the Particle Gibbs step, Algorithm \ref{alg:pgibbs}, offers a very simple way
to change $N_x$ during the course of the algorithm: In Step (2), simply re-generate a particle system
(conditional on selected trajectory $\tilde{x}_{0:t}^1$) of size $N_x^\mathrm{new}$. But, as already discussed, 
such a strategy requires then to access past particle values $x_s^{n}$ (and also $a_s^n)$, rather than only current particle values
$x_t^n$. 

This problem may be addressed in two ways. First, one may remark that, to implement Particle Gibbs, one needs to store only those $x_s^n$ (and 
$a_s^n$) which have descendant among the $N_x$ current particles $x_t^n$. \cite{pathstorage} developed such a path storage approach, 
and gave conditions on the mixing of Markov chain $(x_t)$ under which this approach has memory cost $\OO(t+N_x\log N_x)$ (for a single 
PF with $N_x$ particles, run until time $t$). Thus, an implementation of this approach within \SSQ~would lead to 
a $\OO(\Nt (t+N_x\log N_x))$ memory cost.

A second approach, developed here, exploits the deterministic nature of PRNGs (pseudo-random number generators): 
a sequence $z_0,z_1,\ldots,z_i,\ldots$ of computer-generated random variates is actually a deterministic sequence 
determined by the initial state (seed) of the PRNG. It is sufficient to store that initial state and $z_0$ in order to 
recover any $z_i$ in the future. The trade-off is an increase in CPU cost, as each access to $z_i$ 
require re-computing $z_1,\ldots,z_i$. 

We apply this idea to the variables $(x_{0:t}^{1:N_x,m},a_{1:t}^{1:N_x,m})$.
By close inspection of Algorithm \ref{alg:smc2}, we note that variables in a `time slice' $(x_s^{1:N_x,m},a_s^{1:N_x,m})$, 
$0<s\leq t$ (or $x_0^{1:N_x,m}$ at time 0) are always generated jointly, either during Step (a), or during Step (c). 
In both cases, this time-slice is a deterministic function of the current PRNG state and the previous time slice. 
Thus, one may recover any time slice (when needed) by storing only (i) the PNRG state (immediately before the generation 
of the time slice); and (ii) in which Step (either (a) or (c)) the time slice was generated. 
This reduces the memory cost of \SSQ~from $\OO(t \Nt N_x)$ to $\OO(\Nt(t+N_x))$.

Compared to the path storage approach mentioned above, our PRNG recycling approach has a larger CPU cost, 
a smaller memory cost, and does not require any 
conditions on the mixing properties of process $(x_t)$. 
Note that the CPU cost increase is within a factor of two, because each time a Particle Gibbs update is performed, 
the number of random variables that must be re-generated (i.e. the $x_s^n$ and $a_s^n$ in Algorithm \ref{alg:pgibbs})
roughly equals the number of random variables that are generated for the first time 
(i.e. the $\tilde{x}_s^n$ and $\tilde{a}_s^n$ in Algorithm \ref{alg:pgibbs}). 

\subsection{Nonparametric estimation of $N_x$}
\label{sec:nonparam}

As seen in Algorithm \ref{alg:smc2}, a Particle Gibbs step will be performed each time the ESS goes below some threshold. 
That the ESS is low may indicate that $N_x$ is also too low, and therefore that the variance of the likelihood estimates
$L_t(\theta^m,x_{0:t}^{1:N_x,m},a_{1:t}^{1:N_x,m})$
is too high. Our strategy is to update (each time a Particle Gibbs step is performed) 
the current value of $N_x$ to $N_x^{\mathrm{new}}=\tau/\hat{\sigma}^2$, where $\hat{\sigma}^2$
is some (possibly rough) estimate of the variance of the \emph{log} likelihood estimates. This is motivated by results from \cite{2012arXiv1210.1871D}, 
who also develop some theory that supports choosing $\tau\approx 1$ is optimal (although their optimality results do
not extend straightforwardly to our settings). 

Assume $\Theta\subset \R^d$. To estimate $\sigma^2$, we use backfitting to fit a GAM (generalized additive model) to 
the responses $R^m=\log L_t(\theta^m,x_{0:t}^{1:N_x,m},a_{1:t}^{1:N_x,m})$:
$$ R^m = \alpha +\sum_{j=1}^d f_j(C_j^m) + \varepsilon^m, $$
using as covariates $C_j^m$ the $d$ principal components of the resampled $\theta$-particles. The estimate $\sigma^2$
is then the empirical variance of the residuals. See e.g. Chap. 9 of \cite{hastie2009elements}
for more details on backfitting and GAM modelling. 

We found this strategy to work well, with the caveat that choosing $\tau$ required some trial and error.

\subsection{Additional considerations}

Using Particle Gibbs as our PMCMC move within \SSQ~hast two advantages: (a) it makes it possible to change $N_x$ without changing the weights, as explained above; and (b) it also makes it possible to update the $\theta^m$ according to Gibbs or Metropolis  step that leaves $\theta|x_{0:t},y_{0:t}$ invariant); see Step (3) of Algorithm \ref{alg:pgibbs}. 
For models where sampling from $\theta|x_{0:t},y_{0:t}$ is not convenient, one may instead update $\theta$ through several PMMH steps performed after the Particle Gibbs step. 

\section{Numerical example}

We consider the following stochastic volatility model: $x_0\sim N(\mu,\sigma^2/(1-\rho^2))$,
$x_t-\mu = \rho(x_{t-1}-\mu)+\sigma\epsilon_t,\quad \epsilon_t\sim N(0,1)$
and $y_t|x_t \sim N(0,e^{x_t})$; thus $\theta=(\mu,\rho,\sigma)$, with $\rho\in[-1,1]$, $\sigma>0$. 
We assign independent priors to the components of $\theta$: $\mu\sim N(0,2^2)$, $\rho\sim N(0,1)$ constrained
to $[-1,1]$, and $\sigma^2\sim IG(3,0.5)$. 
The dataset consists in log-returns from the monthly SP500 index, observed from 29/05/2013 to 19/12/2014; $T=401$. 

\begin{figure}
\begin{center}
\includegraphics[width=8.4cm]{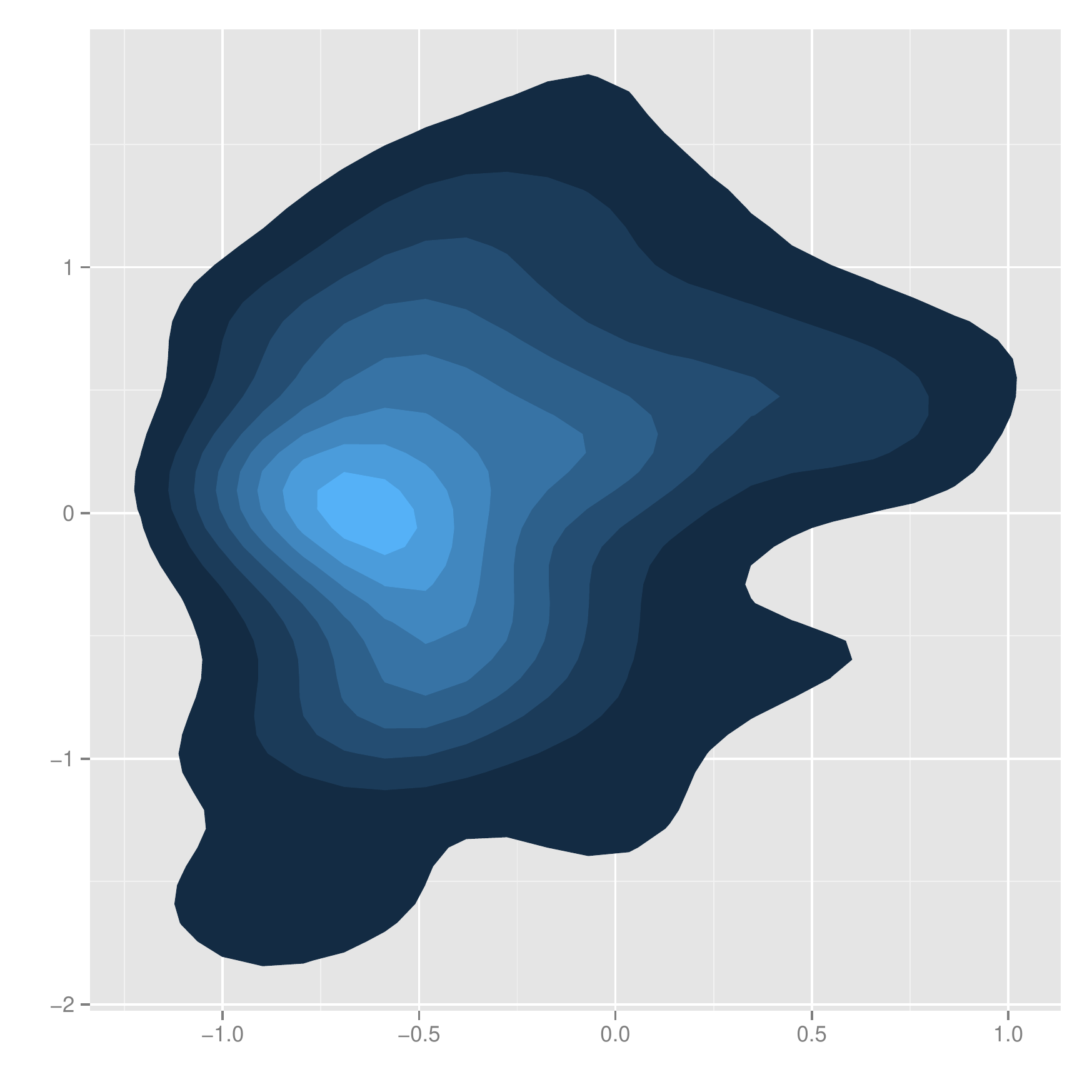}    
\caption{Marginal posterior $p(\sigma^2,\rho|y_{0:15})$, as approximated by \SSQ~run until $t=15$, 
and linearly transformed so that axes are the two principal components.} 
\label{fig:backfitting}
\end{center}
\end{figure}

Figure \ref{fig:backfitting} plots the marginal posterior $p(\rho,\sigma^2|y_{0:15})$, as approximated by 
\SSQ, run up to time $15$. This figure illustrates the need for modelling nonparametrically the true likelihood as 
a function of $\theta$, in order to estimate the variance of the estimated likelihood. 

For this model, sampling jointly from $\theta|x_{0:t},y_{0:t}$ is 
difficult, but it is easy to perform a Gibbs step that leaves invariant $\theta|x_{0:t},y_{0:t}$, as 
the full conditionals of each component (e.g. $\mu|\sigma,\rho,x_{0:t},y_{0:t}$ and so on) 
are standard distributions. Let's call `full PG' Algorithm  \ref{alg:pgibbs}, where Step 2 
consists of this Gibbs step for $\theta|x_{0:t},y_{0:t}$; and conversely let's call `partial PG' Algorithm  \ref{alg:pgibbs}
with $\tilde{\theta}=\theta$ in Step 2 ($\theta$ is not updated). 

We compare four versions of \SSQ: 
(a) the standard version, as proposed in \cite{smc2} (i.e. Step (c) of Algorithm \ref{alg:smc2} is a PMMH step, 
and that step is followed by an exchange step 
to double $N_x$ when the acceptance rate of PMMH is below $20\%$); 
(b) the same algorithm, except that an exchange step is systematically performed after Step (c), and 
$N_x$ is set to the value obtained with our non-parametric approach (see Section \ref{sec:nonparam});
(c) the version developed in this paper, with full PG steps (and $N_x$ updated through the non-parametric procedure);
(d) the same algorithm, but with partial PG steps, followed by 3 PMMH steps to update $\theta$. 

The point of Algorithm (b) is to show that adapting $N_x$ too often during the course of the algorithm
is not desirable when using the exchange step, as this leads to too much variance. 
The point of Algorithm (d) is to see how our approach performs when 
sampling from $\theta|x_{0:t},y_{0:t}$ (either independently or through MCMC) is not feasible.

\begin{figure}
\begin{center}
\includegraphics[width=8.4cm]{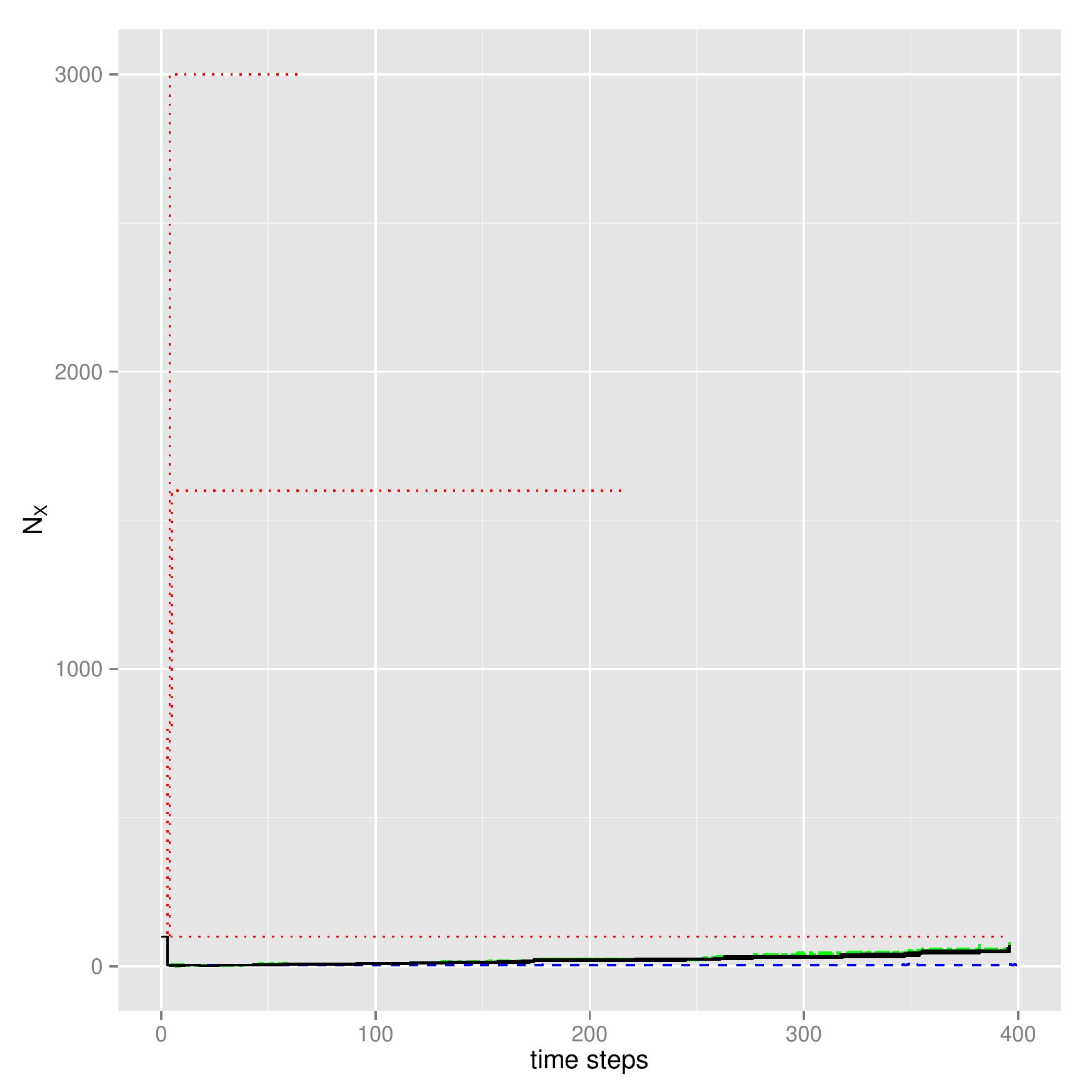}    
\caption{Evolution of $N_x$ over time for 5 runs of the four considered \SSQ~algorithms; 
red dotted line is Algorithm (a), blue dashed is (b), black solid is (c), green double-dashed is (d).
Results of (c) and (d) are nearly undistinguishable. 
} 
\label{fig:Nx}
\end{center}
\end{figure}

Figure \ref{fig:Nx} plots the evolution of $N_x$ over time for the four \SSQ~algorithms. One sees that, for these model and dataset, 
the CPU cost of the standard \SSQ~algorithm is quite volatile, as $N_x$ increases very quickly in certain runs. 
In fact certain runs are incomplete, as they were stopped when the CPU time exceeded $10$ hours. 
On the other hand, the CPU cost of other versions  is more stable across runs, and, more importantly, quite lower. 

\begin{figure}
\begin{center}
\includegraphics[width=8.4cm]{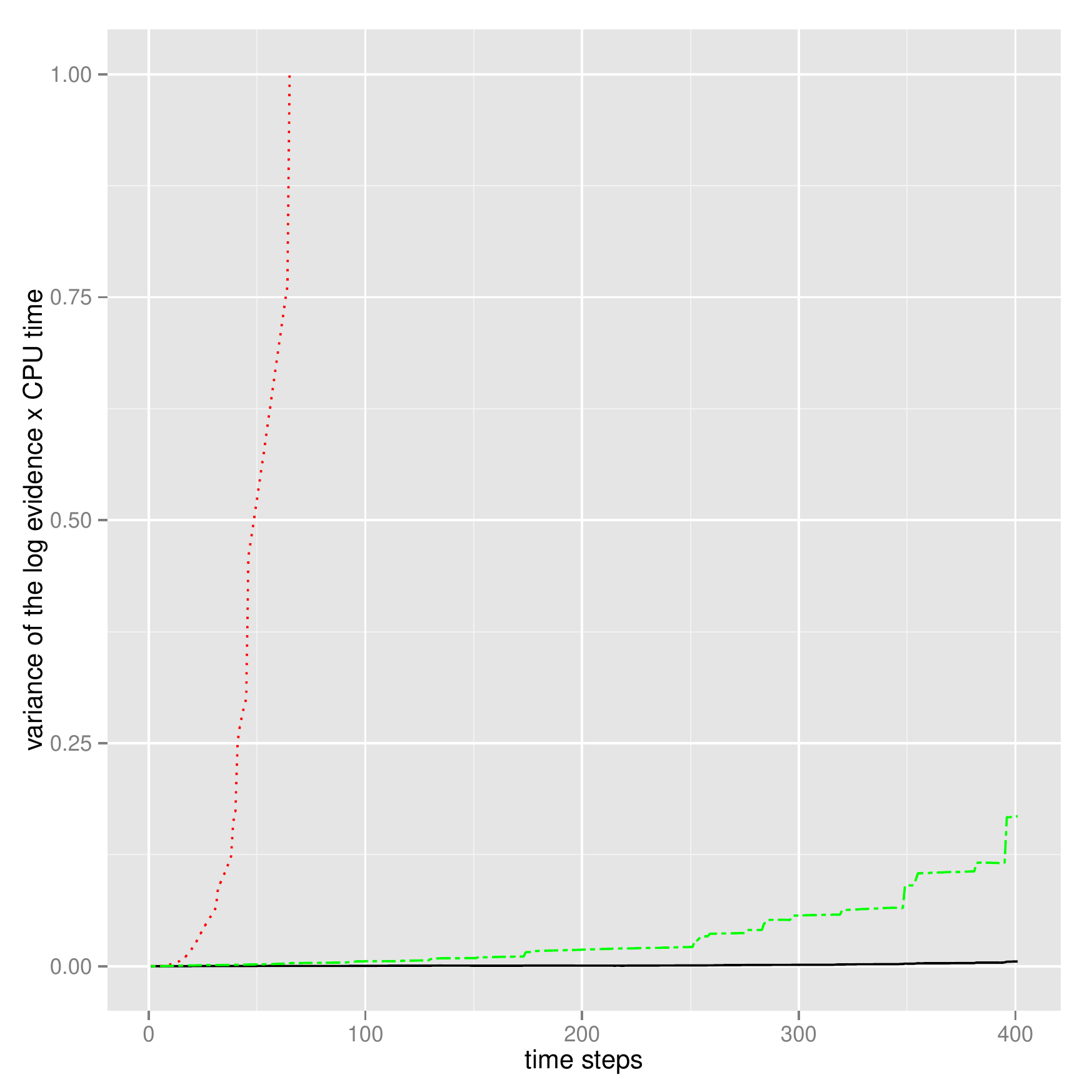}    
\caption{Empirical variance of estimated marginal likelihood 
$p(y_{0:t})$ multiplied by average CPU time; same legend as 
Figure \ref{fig:Nx}, results from Algorithm (b) are omitted.} 
\label{fig:var}
\end{center}
\end{figure}

Figure \ref{fig:var} plots the empirical variance of the estimated marginal likelihood (evidence, $p(y_{0:t})$), normalised 
with the running time up to time step $t$. 
One observes that version (c) does quite better than (d), and far much better than (a). Results from Algorithm (b) were
to variable to be included.

\begin{figure}
\begin{center}
\includegraphics[width=8.4cm]{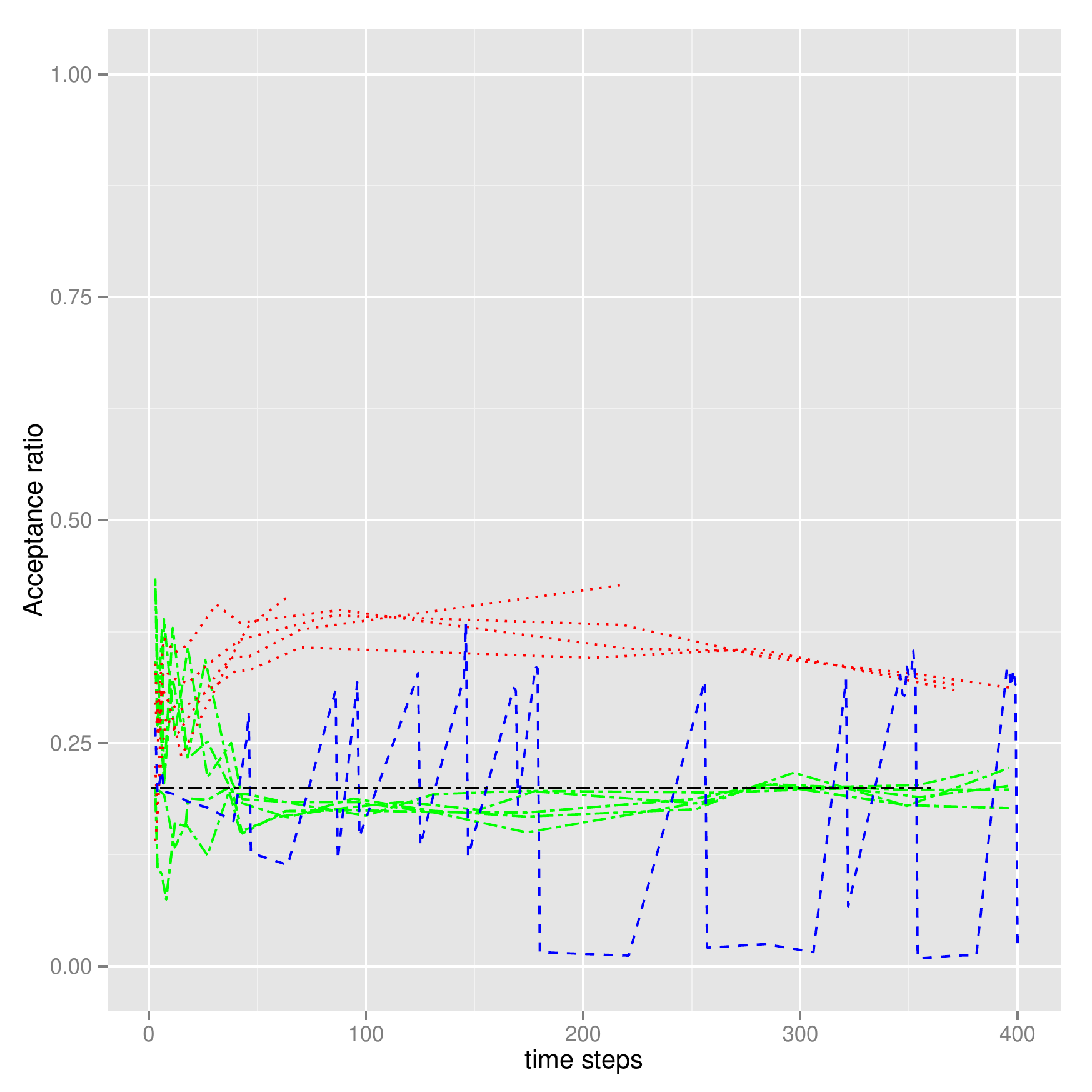}    
\caption{PMMH acceptance rate across time; same legend as Figure \ref{fig:Nx}.
Black line marks $20\%$ target.
} 
\label{fig:acc}
\end{center}
\end{figure}

\begin{figure}
\begin{center}
\includegraphics[width=8.4cm]{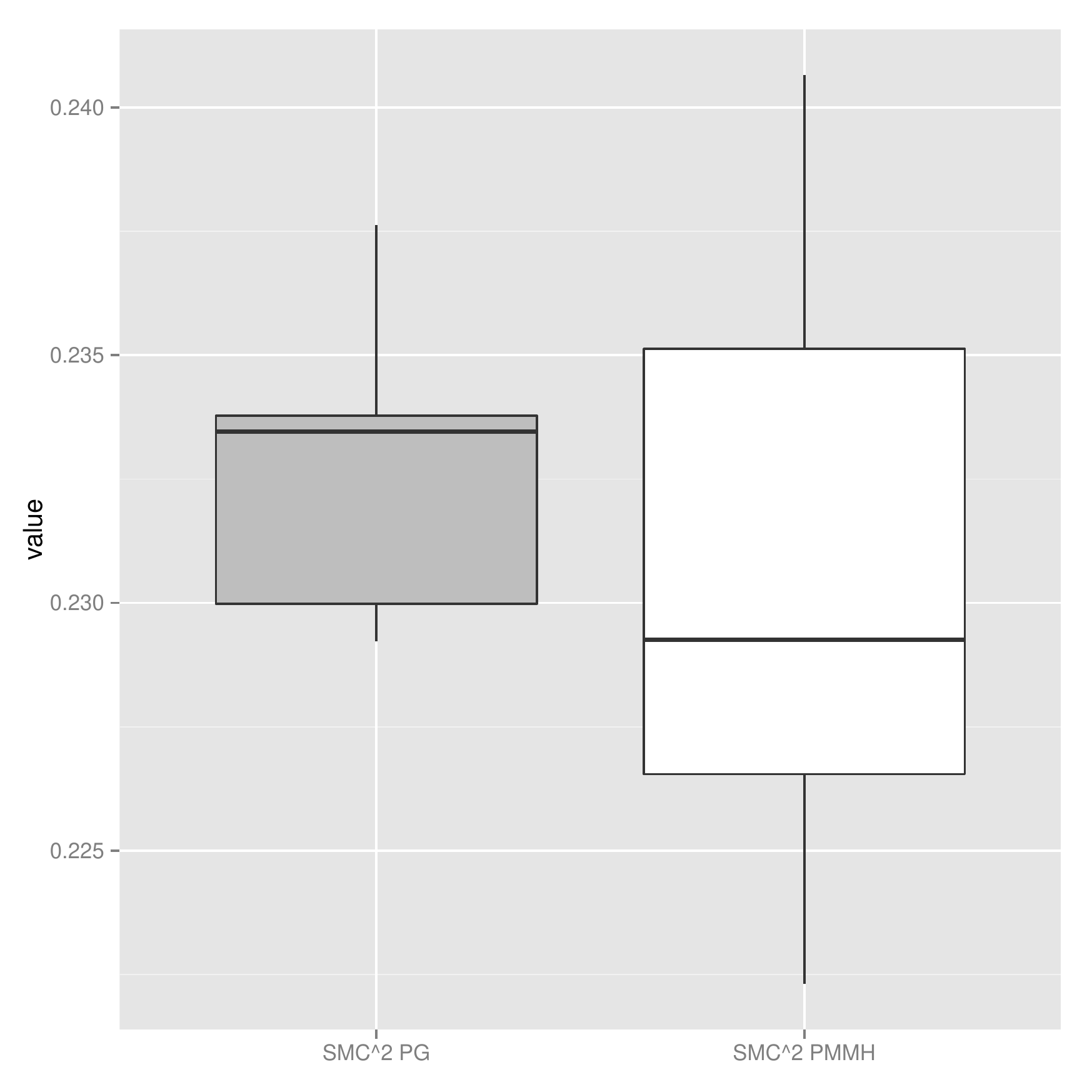}    
\caption{Box-plots of posterior estimate of parameter $\sigma$ at final time $T$, over repeated runs of Algorithm (c)
(left panel) and Algorithm (d) (right panel).  
} 
\label{fig:boxplots}
\end{center}
\end{figure}

Figure \ref{fig:acc} plots the acceptance rate of PMMH steps for Algorithms (a), (b) and (d). (Recall that Algorithm (c) does not perform
PMMH steps). Note the poor performance of Algorithm (b). 
Figure \ref{fig:boxplots} compares the box-plots of posterior estimates of $\sigma$ at final time $T$, 
obtained from several runs of Algorithms (c) and (d). 
Algorithm (c) shows slightly less variability, while being $30\%$ faster on average. 
One sees that the improvement brought by ability to sample from $\theta|x_{0:t},y_{0:t}$ is modest 
here for parameter estimation, but recall that in Figure \ref{fig:var}, the improvement was more substantial.

%
%

\begin{ack}
We thank Pierre Jacob for useful comments. 
\end{ack}

\bibliography{complete}   

\begin{thebibliography}{8}
\providecommand{\natexlab}[1]{#1}
\providecommand{\url}[1]{\texttt{#1}}
\providecommand{\urlprefix}{URL }
\expandafter\ifx\csname urlstyle\endcsname\relax
  \providecommand{\doi}[1]{doi:\discretionary{}{}{}#1}\else
  \providecommand{\doi}{doi:\discretionary{}{}{}\begingroup
  \urlstyle{rm}\Url}\fi

\bibitem[{Andrieu et~al.(2010)Andrieu, Doucet, and Holenstein}]{pmcmc}
Andrieu, C., Doucet, A., and Holenstein, R. (2010).
\newblock Particle {M}arkov chain {M}onte {C}arlo methods.
\newblock \emph{J. R. Statist. Soc. B}, 72(3), 269--342.
\newblock \doi{10.1111/j.1467-9868.2009.00736.x}.

\bibitem[{{Andrieu} et~al.(2013){Andrieu}, {Lee}, and
  {Vihola}}]{AndrieuLeeVihola}
{Andrieu}, C., {Lee}, A., and {Vihola}, M. (2013).
\newblock {Uniform Ergodicity of the Iterated Conditional SMC and Geometric
  Ergodicity of Particle Gibbs samplers}.
\newblock \emph{ArXiv e-prints}.

\bibitem[{Chopin(2002)}]{Chopin:IBIS}
Chopin, N. (2002).
\newblock A sequential particle filter for static models.
\newblock \emph{Biometrika}, 89, 539--552.

\bibitem[{Chopin et~al.(2013)Chopin, Jacob, and Papaspiliopoulos}]{smc2}
Chopin, N., Jacob, P., and Papaspiliopoulos, O. (2013).
\newblock {SMC$^2$: A sequential Monte Carlo algorithm with particle Markov
  chain Monte Carlo updates}.
\newblock \emph{J. R. Statist. Soc. B}, 75(3), 397--426.

\bibitem[{Del~Moral(1996)}]{DelMoral1996unbiased}
Del~Moral, P. (1996).
\newblock Non-linear filtering: interacting particle resolution.
\newblock \emph{Markov processes and related fields}, 2(4), 555--581.

\bibitem[{{Doucet} et~al.(2012){Doucet}, {Pitt}, {Deligiannidis}, and
  {Kohn}}]{2012arXiv1210.1871D}
{Doucet}, A., {Pitt}, M., {Deligiannidis}, G., and {Kohn}, R. (2012).
\newblock {Efficient implementation of Markov chain Monte Carlo when using an
  unbiased likelihood estimator}.
\newblock \emph{ArXiv preprint}.

\bibitem[{Hastie et~al.(2009)Hastie, Tibshirani, Friedman, Hastie, Friedman,
  and Tibshirani}]{hastie2009elements}
Hastie, T., Tibshirani, R., Friedman, J., Hastie, T., Friedman, J., and
  Tibshirani, R. (2009).
\newblock \emph{The elements of statistical learning}, volume~2.
\newblock Springer.

\bibitem[{Jacob et~al.(2013)Jacob, Murray, and Rubenthaler}]{pathstorage}
Jacob, P., Murray, L., and Rubenthaler, S. (2013).
\newblock Path storage in the particle filter.
\newblock \emph{Statist. Comput.}, 1--10.
\newblock \doi{10.1007/s11222-013-9445-x}.
\newblock \urlprefix\url{http://dx.doi.org/10.1007/s11222-013-9445-x}.

\end{thebibliography}

\end{document}